# Reproduction of High-frequency Vibration Experience by a Sensory Equivalent Conversion Method for the Standard Haptic SDK of Meta Quest 3

Daito IGARASHI[1], Masashi KONYO[1], and Satoshi TADOKORO[1]

[1] Graduate School of Information Sciences. Tohoku University, Aramaki Aza Aoba, Aoba, Sendai, Miyagi, 980-8579, Japan

(Email:konyo@rm.is.tohoku.ac.jp)

**Abstract** --- In recent years, there has been a growing demand for realistic haptic experiences in VR. This study employs the Intensity Segment Modulation (ISM) method to convert high-frequency vibrations into low-frequency amplitude-modulated waves on VR devices. We verified the generation of high-immersion vibration stimuli synchronized with audio. Additionally, we compared the perception of ISM-converted vibrations between VR devices and other devices to evaluate the effectiveness of ISM in VR environments. A demonstration on the Meta Quest 3 was also developed, allowing users to experience the tactile sensation of handheld fireworks.

## 1 Introduction

In recent years, the need to enhance the realism of haptic feedback in VR devices has increased. Utilizing acoustic signals, including high-frequency vibrations, for haptic feedback is expected to create more realistic and immersive experiences. However, standard SDKs cannot present vibrations exceeding several hundred hertz due to actuator response limits and noise issues.

In this study, we employ Intensity Segment Modulation (ISM) [1] to convert vibrations containing high frequencies into low-frequency amplitude-modulated waves. This method preserves the perception of the original high frequencies while converting them into signals playable by standard haptic SDKs. Hoshi et al. have confirmed the effectiveness of a similar approach applied to the iPhone API [2]. This study attempts to apply ISM to VR device controllers.

Additionally, we conducted subjective evaluations of the perceived vibrations when high-frequency vibrations converted by ISM were played back as low-frequency vibrations using the standard SDK. This allowed us to assess whether the original high-frequency perception could be maintained and whether high-fidelity haptics could be reproduced in VR devices.

## 2 Intensity Segment Modulation Overview

ISM [1] is a conversion method based on the intensity perception characteristics of Pacinian corpuscles for high-frequency vibrations, as defined by Equation (1), and envelope perception characteristics[4][5].

It calculates the intensity of the original audio source for each segment and maintains intensity fluctuations of approximately 100 Hz or lower, allowing the high-frequency tactile sensation to be preserved while converting it to a single amplitude-modulated wave of a relatively low frequency, such as 150–300 Hz.

$$I(f) = \left[\left(\frac{A_T(f)}{A(f)}\right)^2\right]^{\alpha(f)} \qquad (1)$$

where $A(f)$ is the amplitude of the vibration at frequency $f$, $A_T(f)$ is the amplitude of the vibration detection threshold at $f$, and $\alpha(f)$ is the frequency-dependent exponent parameter. In ISM, by controlling the frequency and amplitude, it is possible to present vibrations with the same perceived intensity as the original audio signal through amplitude-modulated waves.

The carrier frequency of the amplitude-modulated can be selected according to the frequency response of the actuator, allowing the conversion of high-frequency tactile sensations into vibration waveforms that the actuator can reproduce.

In VR device controllers, LRA-type actuators with limited bandwidth are commonly used. By employing ISM, it is expected that the high-frequency tactile sensations can be accurately reproduced.

### 2.1 Conversion procedure by ISM

The signal segmented in the time domain is filtered in response to the input signal. High-frequency signals above 100 Hz are decomposed by frequency, and their perceptual intensities are summed. During playback, the amplitude is determined to reproduce the perceptual intensity according to the carrier frequency of the amplitude-modulated wave being used.

## 3 VR DEVICES USED AND THEIR HAPTIC SYSTEMS

In this study, we use Meta's Meta Quest 3 and Meta Quest Touch Plus controllers as the VR environment. These devices were selected for their high-quality built-in actuators and the availability of the Meta XR Haptics SDK, which provides high-quality haptic feedback.

However, this SDK does not allow direct specification of vibration waveforms. Therefore, it is necessary to generate amplitude-modulated waves based on the parameters available in the SDK. The parameters that can be specified in the haptic file handled by the SDK include Amplitude, which corresponds to the vibration intensity, and Frequency, which corresponds to the vibration frequency.

To represent ISM vibrations on the Meta Quest 3, the intensity series of the audio source is converted into an amplitude series. Using predetermined parameters and their relationship to vibration amplitude, a haptic file compatible with the Meta Quest 3 was created for presentation.

## 4 EVALUATION OF TACTILE CHANGES IN VR DEVICES

### 4.1 experimental procedure

We investigate the difference in tactile sensations between ISM-generated vibrations outputted by VR devices and those outputted by an existing high-response actuator (Foster Electric ACTUATOR 639897) from high-frequency audio sources.

In the VR device setup, vibrations were presented using Unity, where the sound and vibrations were activated while the middle finger button of the controller was pressed.

For the existing actuator setup, we prepared a stereo audio source containing both the original audio information and the ISM-converted information, inputting each into the earphones and the actuator, respectively. To approximate the feeling of the controller's vibrations, the actuator was encased in a resin case fabricated by a 3D printer, and participants were instructed to hold the actuator in their hand.

We adjusted the perceived intensity of vibrations and audio for both the actuator and the VR device using a 150 Hz sine wave to ensure similar perceived strength.

The vibration stimuli included high-frequency audio sources such as the sound of a saw cutting wood (saw), the sound of handheld fireworks (fireWorks), the sound of glass breaking (glassCrash), and the sound of tape ripping (tapeRip). The vibrations were converted using ISM and presented along with the corresponding audio through earphones in each condition.

Participants compared the tactile sensations of ISM-generated vibrations on the VR device to those of the external actuator using a 7-point Likert scale (1: very poor, 2: poor, 3: slightly poor, 4: equivalent, 5: slightly good, 6: good, 7: very good), with the option to specify to the first decimal place. The participants were six males in their twenties.

### 4.2 experimental results

The experimental results are shown in Fig.1. The median ratings by participants were around 4: equivalent for all vibrations, indicating that the VR controller could present tactile sensations comparable to those of high-response actuators. However, there were variations in responses depending on the type of audio source.

For the saw and tape sounds, the ratings showed less variability, with participants' evaluations being consistent. The saw's rating was slightly lower at 3: somewhat poor. This could be due to a possible desynchronization between the audio source and the vibrations generated through Unity on the VR device. The saw audio had periodic increases in volume corresponding to the sawing motion, and the desynchronization that was not perceived in other audio sources might have impacted the evaluation.

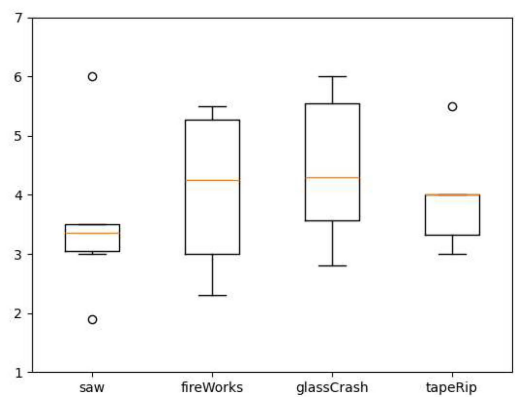

Fig. 1 Evaluation of each vibration presentation

For the handheld fireworks and glass breaking sounds, the median ratings exceeded 4: equivalent, but there was more variability in the responses. Factors contributing to higher ratings included the controller's fit in the hand, which made it easier to perceive even small vibrations. On the other hand, factors leading to lower ratings might include the controller's lower damping performance, which could leave residual vibrations and cause the vibrations to feel less sharp in response to the sudden changes characteristic of fireworks and glass breaking sounds.

## 5 DEMONSTRATION

In this demonstration, participants will compare the vibrations converted on MetaQuest3 using ISM with those converted using Meta Haptics Studio, a sound-to-vibration conversion software provided by Meta.

Participants will experience the difference in the converted vibrations of various sound sources, such as the sound of hand-held fireworks and popping soap bubbles, using each method.

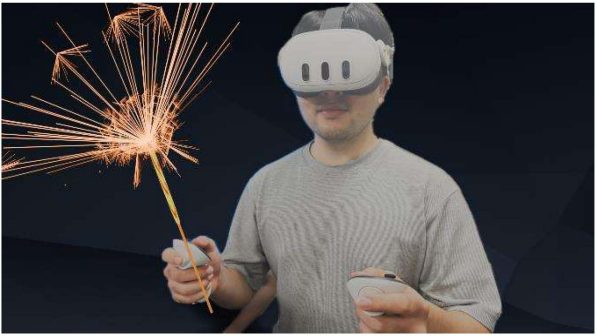

**Virtual reality**

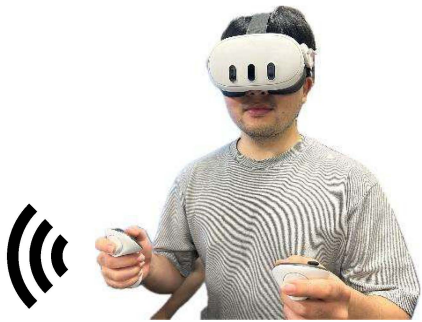

**Real**

**Fig. 2 Demonstration Overview**

## 6 CONCLUSION

In this study, we proposed the use of perceptual equivalence conversion through Intensity Segment Modulation (ISM) as a method to reproduce high-frequency vibrations on general-purpose VR devices. We investigated the vibration characteristics of the Meta Quest 3 controller using its standard haptics SDK and successfully replicated the tactile sensation to a level comparable to high-response actuators by applying ISM.

However, VR devices tend to feel less responsive to sudden changes, and desynchronization between audio and vibrations could impact the tactile experience.